%% file: pamela_isotopes.tex
\documentclass[iop]{emulateapj}
\usepackage{amssymb,amsmath,xcolor}
\newcommand{\pam}{\textsf{PAMELA}}
\newcommand{\deu}{\textsuperscript{2}H}
\newcommand{\prot}{\textsuperscript{1}H}
\newcommand{\het}{\textsuperscript{3}He}
\newcommand{\hef}{\textsuperscript{4}He}

\submitted{}

\begin{document}

\title{Measurement of the isotopic composition of hydrogen and helium nuclei in cosmic rays with the \pam\ experiment}

\author
{O. Adriani$^{1,2}$, G. C. Barbarino$^{3,4}$, G. A. Bazilevskaya$^{5}$, R. Bellotti$^{6,7}$,
M. Boezio$^{8}$, \\
E. A. Bogomolov$^{9}$, M. Bongi$^{1,2}$, V. Bonvicini$^{8}$,
S. Borisov$^{10,11,12}$, S. Bottai$^{2}$, \\
A. Bruno$^{6,7}$, F. Cafagna$^{7}$, D. Campana$^{4}$,
R. Carbone$^{8}$, P. Carlson$^{13}$, \\
M. Casolino$^{10,14}$, G. Castellini$^{15}$, I. A. Danilchenko$^{12}$
M. P. De Pascale$^{10,11,\dagger}$, C. De Santis$^{11}$, \\N. De Simone$^{11}$,
V. Di Felice$^{11}$, V. Formato$^{8,16}$, 
 A. M. Galper$^{12}$, %L. Grishantseva$^{12}$,
A. V. Karelin$^{12}$, \\
S. V. Koldashov$^{12}$, S. Koldobskiy$^{12}$, S. Y. Krutkov$^{9}$, A. N. Kvashnin$^{5}$,
 A. Leonov$^{12}$,\\
V. Malakhov$^{12}$,
L. Marcelli$^{11}$, 
A. G. Mayorov$^{12}$, W. Menn$^{17}$, V. V. Mikhailov$^{12}$,\\
E. Mocchiutti$^{8}$,
A. Monaco$^{6,7}$,  N. Mori$^{2}$, N. Nikonov$^{9,10,11}$, G. Osteria$^{4}$,
F. Palma$^{10,11}$, \\
P. Papini$^{2}$, M. Pearce$^{13}$, P. Picozza$^{10,11}$, 
C. Pizzolotto$^{8,18,19}$ M. Ricci$^{20}$,
S. B. Ricciarini$^{15}$, \\
L. Rossetto$^{13}$, R. Sarkar$^{8}$, M. Simon$^{17}$,
 R. Sparvoli$^{10,11}$, P. Spillantini$^{1,2}$, \\
Y. I. Stozhkov$^{5}$, A. Vacchi$^{8}$, 
E. Vannuccini$^{2}$, G. Vasilyev$^{9}$, S. A. Voronov$^{12}$, \\
 Y. T. Yurkin$^{12}$, J. Wu$^{13,*}$,
 G. Zampa$^{8}$, N. Zampa$^{8}$, 
 V. G. Zverev$^{12}$, \\
}

\affil{$^{1}$University of Florence, Department of Physics, I-50019 Sesto Fiorentino, Florence, Italy}
\affil{$^{2}$INFN, Sezione di Florence, I-50019 Sesto Fiorentino, Florence, Italy}
\affil{$^{3}$University of Naples ``Federico II'', Department of Physics, I-80126 Naples, Italy}
\affil{$^{4}$INFN, Sezione di Naples,  I-80126 Naples, Italy}
\affil{$^{5}$Lebedev Physical Institute, RU-119991, Moscow, Russia}
\affil{$^{6}$University of Bari, Department of Physics, I-70126 Bari, Italy}
\affil{$^{7}$INFN, Sezione di Bari, I-70126 Bari, Italy}
\affil{$^{8}$INFN, Sezione di Trieste, I-34149 Trieste, Italy}
\affil{$^{9}$Ioffe Physical Technical Institute,  RU-194021 St. Petersburg, Russia}
\affil{$^{10}$INFN, Sezione di Rome ``Tor Vergata'', I-00133 Rome, Italy}
\affil{$^{11}$University of Rome ``Tor Vergata'', Department of Physics,  I-00133 Rome, Italy}
\affil{$^{12}$National Research Nuclear University MEPhI, RU-115409 Moscow}
\affil{$^{13}$KTH, Department of Physics, and the Oskar Klein Centre for Cosmoparticle Physics, AlbaNova University Centre, SE-10691 Stockholm, Sweden}
\affil{$^{14}$RIKEN, Advanced Science Institute, Wako-shi, Saitama, Japan}
\affil{$^{15}$IFAC, I-50019 Sesto Fiorentino, Florence, Italy}
\affil{$^{16}$University of Trieste, Department of Physics, I-34147 Trieste, Italy}
\affil{$^{17}$Universit\"{a}t Siegen, Department of Physics, D-57068 Siegen, Germany}
\affil{$^{18}$INFN, Sezione di Perugia, I-06123 Perugia, Italy}
\affil{$^{19}$Agenzia Spaziale Italiana (ASI) Science Data Center, I-00044 Frascati, Italy}
\affil{$^{20}$INFN, Laboratori Nazionali di Frascati, Via Enrico Fermi 40, I-00044 Frascati, Italy}
\affil{$^{*}$On leave from  School of Mathematics and Physics, China University of Geosciences, CN-430074 Wuhan, China}
\affil{$^{\dagger}$Deceased}

\begin{abstract}
{%\color{blue}
%The cosmic-ray hydrogen and helium (\prot,\deu,\het,\hef)
%isotopic composition between 100 MeV/n and 1 GeV/n is presented
%for the 23rd solar minimum from July 2006 to the end of 2007.
%The results were obtained with the \pam\ experiment, which was launched into low-Earth
%orbit on-board the Resurs-DK1 satellite on June 15\textsuperscript{th} 2006. Velocity and rigidity
%information allows isotopes for $Z = 1$ and $Z = 2$ particles to be identified with unprecedented statistics.}
The satellite-borne experiment \pam\ has been used to make new measurements of cosmic ray H and He isotopes.
The isotopic composition was measured between 100 and 600 MeV/n for hydrogen and between 100 and 900 MeV/n for helium isotopes
over the 23\textsuperscript{rd} solar minimum from July 2006 to December 2007.
The energy spectrum of these components carries fundamental information regarding
the propagation of cosmic rays in the galaxy which are competitive with those obtained
from other secondary to primary measurements such as B/C.
}
\end{abstract}

\keywords{Astroparticle physics, cosmic rays}

\section{Introduction}

%hydrogen and helium isotopes in cosmic rays are generally believed to
%be of secondary origin, resulting from the nuclear  
%interactions of primary cosmic-rays (GCRs) with the
%interstellar medium, mainly through spallation of primary  
%\hef\ nuclei or through the resonant scattering $p+p \rightarrow
%{}^2\text{H}+\pi^+$. The spectral shape of the secondary isotopes is therefore 
%completely determined by the source spectrum of the parent elements
%and by the propagation process. Measurements of the secondary isotopes
%spectra are then a powerful tool to costrain the parameters of the
%galactic propagation models.
%This can be crucial when analyzing positron and antiproton spectra, which is
%the primary goal of several current cosmic-ray experiments. 
%\citep{1989AdSpR...9..145S,2002ApJ...564..244W}.

Hydrogen and helium isotopes in cosmic rays are generally believed to
be of secondary origin, resulting from the nuclear  
interactions of primary cosmic-ray protons and \hef\ with the
interstellar medium, mainly through spallation of primary  
\hef\ nuclei or through the reaction $p+p \rightarrow \text{ }^2\text{H}+\pi^+$.
%Thus, these isotopes are tracers of GCR transport in the Galaxy and
%can be used to 
%constrain propagation parameters.  
These isotopes can be used to study and constrain 
parameters in propagation models for galactic cosmic rays (GCRs)
\citep{2007ARNPS..57..285S,2012Ap&SS.342..131T,2012A&A...539A..88C}. 
$^2$H and $^3$He are the most abundant secondary isotopes in 
galactic cosmic-rays and have peculiar features: 
 $^2$H is the only secondary species 
(apart from antiprotons) that can also be produced in proton-proton
interactions and 
$^3$He is the only secondary fragment with a $A/Z$ 
significantly different from two ($A$ and $Z$ being respectively the mass and charge number).  

The importance of light isotopes has been known for about 40 years, when
the first measurements became available \citep{1975ICRC....1..319G,1975ApJ...202..265G,1976ApJ...206..616M,1978ApJ...221.1110L}.  
Measurements require very good mass resolution, a challenge for instruments
deployed in space. 
With the exception of the results from AMS-01 \citep{2011ApJ...736..105A_red} 
most of the measurements were performed using stratospheric balloons
\citep{2002ApJ...564..244W,1998ApJ...496..490R_red,1995ICRC....2..630W,1991ApJ...380..230W,1993ApJ...413..268B_red}, 
where the residual atmosphere above the instrument caused a
non-negligible background of secondary particles. The
atmospheric background estimation is subject to large 
uncertainties (e.g. the limited knowledge of
isotope production cross sections). 
Due to these limitations, experimental errors are generally very large
and the focus of measurements therefore shifted
to other secondary species, like boron or sub-iron nuclei \citep{1998ApJ...509..212S}.   

%\color{blue} 
The light-isotope quartet offers an independent unique way to address
the issue of ``universality in GCR propagation'',
which can be crucial when analysing antiproton and
positron spectra to 
search for possible primary signals (see e.g.~\citet{1989AdSpR...9..145S}). 
%{Since 2006, the \pam\ experiment has been measuring both hydrogen and helium isotopes over a long 
%period of time with reduced instrumental uncertainty and with no
%environmental systematics, such as the residual atmosphere 
%present in balloon experiments.}
{%\color{red}
The \pam\ experiment has been observing GCRs over the 23\textsuperscript{rd} solar miminum
since July 2006 at an altitude ranging from $\sim 350$ km to $\sim 600$ km on-board
of the Russian Resurs-DK1 satellite which executes a quasi-polar orbit.
The low-earth orbit allows \pam\ to perform measurements in an environment
free from the background induced by atmospheric cosmic rays.
%while the $70^\circ$ inclination allows \pam\ to study particles down to the minimum detectable rigidity
}

%In this paper the isotopic composition of hydrogen and helium in the energy range 
%from 0.1 to about 1 GeV nucleon\textsuperscript{-1} (corresponding to rigidities from $\sim800$ 
%MV to $\sim 2.5$ GV) obtained from the \pam\ data collected until December 2007 is presented.
The results presented here are based on the data set
collected by \pam\ between
July 2006 and December 2007. From about $10^9$
triggered events, accumulated during a total acquisition
time of 528 days, 5,378,795 hydrogen nuclei were
selected in the energy interval between 100 and 600 MeV/n and 1,749,964 helium nuclei between 100 and 900 MeV/n.
The data presented here replace and complete the preliminary results presented in \cite{2011ASTRA...7..465C}.
A more complete evaluation of the selection efficiencies and contamination due to an improved simulation
resulted in better reconstructed \deu\ and \het\ spectra and \het/\hef\ ratio.

\section{The \pam\ apparatus}

%In this section we  describe   the main characteristics of \pam\
%detector; a more detailed description of the device and the data
%handling  can be found in \citep{Pi07, cpu, yoda}. 
The \pam\ spectrometer was designed and built to study the antimatter component  
of cosmic rays from tens of MeV up to hundreds of GeV and with a significant
increase in statistics with respect to previous experiments.
To reach this goal the apparatus was optimized for the study of $Z=1$ particles and to reach a high 
level of electron-proton discrimination. 
\begin{figure}[t]
    \centering
%    \epsscale{0.75}
    \plotone{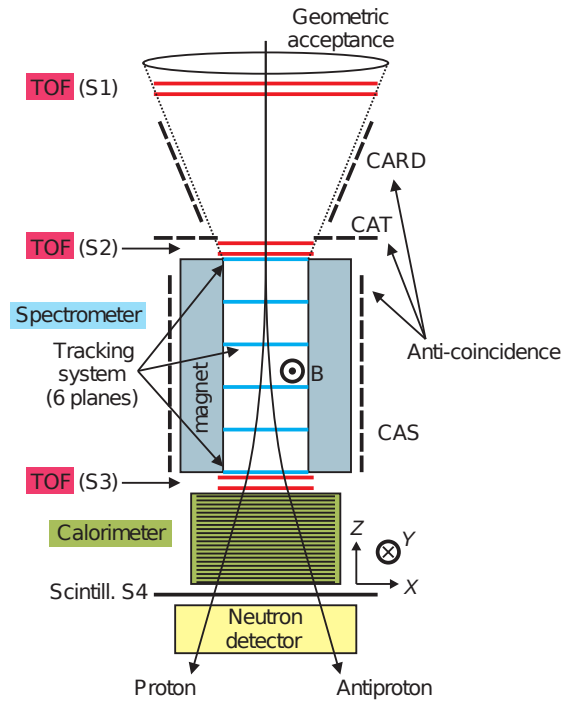}
    \caption{A schematic overview of the \pam\ satellite experiment. The experiment
      stands $\sim$1.3 m high and, from top to bottom, consists of a time-of-flight (ToF)
  	  system (S1, S2, S3 scintillator planes), an anticoincidence shield system, a
      permanent magnet spectrometer (the magnetic field runs in the $y$-direction),
      a silicon-tungsten electromagnetic calorimeter, a shower tail scintillator (S4), and a
      neutron detector.}
    \label{im:pamela}
\end{figure}

The core of the instrument (Fig. \ref{im:pamela}) is a permanent magnet with an almost uniform magnetic field
inside the magnetic cavity which houses six planes of double-sided silicon microstrip detectors to measure the
trajectory of incoming particles. The spatial resolution is $\sim3$ $\mu$m in the bending view (also referred
as $x$-view) and $\sim11$ $\mu$m in the non-bending view (also referred as $y$-view). The main task 
of the magnetic spectrometer is to measure particle rigidity $\rho = pc/Ze$ ($p$ and 
$Ze$ being respectively the particle momentum and charge, and $c$ the speed of light) and ionization
energy losses ($dE/dx$).

The Time-of-Flight (ToF) system comprises three double layers of plastic scintillator paddles (S1, S2, and S3,
as shown in Fig \ref{im:pamela}) with the first two placed above and the third immediately below 
the magnetic spectrometer. The ToF system provides 12 independent measurements of the particle velocity,
$\beta = v/c$, combining the time of passage information with the track length derived from the magnetic 
spectrometer. By measuring the particle velocity the ToF system discriminates between down-going particles
and up-going splash albedo particles thus enabling the spectrometer to establish the sign of the particle 
charge. The ToF system also provides 6 independent $dE/dx$ measurements, one for each scintillator plane.

A silicon-tungsten electromagnetic sampling calorimeter made of 44 single-sided silicon microstrip 
detectors interleaved with 22 plates
of tungsten absorber (for a total of 16.3 $X_0$) mounted below the spectrometer is used for 
hadron/lepton separation with a shower tail catcher scintillator (S4). A neutron detector at the 
bottom of the apparatus helps to increase this separation. 

The anticoincidence (AC) system comprises 4 scintillators surrounding the magnet (CAS), one surrounding
the cavity entrance (CAT) and 4 scintillators surrounding the volume between S1 and S2 (CARD).
The system is used to reject events where the presence of secondary particles 
generates a false trigger or the primary particle suffers an inelastic interaction. 

The readout electronics, the interfaces with the CPU and
all primary and secondary power supplies are housed around the detectors. 
%All systems (power
%supply, readout boards etc.)  are redundant with the exception of
%the CPU which is more tolerant to failures. 
The apparatus is enclosed in a pressurized container  
attached to the side of the Resurs-DK1 satellite.  The total weight of \pam\ is 470 kg while the power 
consumption is 355 W. A more detailed description of the instruments and the data
handling  can be found in \cite{Pi07}.

\section{Data analysis}
\subsection{Event selection}
%This analysis was carried out on the data-set collected by \pam\ between July 2006 and December 2007, discarding 
The month of December 2006 was discarded to avoid possible biases from the solar particle events that
took place during the 13\textsuperscript{th} and 14\textsuperscript{th} of December.
%{\color{red} 
%The $70^\circ$ inclination of the satellite orbit allows \pam\ to study particles down to the .
%}
{%\color{red}
The event selections adopted are similar to those used in previous works on the high energy proton and 
helium fluxes \citep{2011Sci...332...69A_red} and on the time dependence of the low energy proton flux \citep{2013ApJ...765...91A}. 
}

\subsubsection{Event quality selections}
In order to ensure a reliable event reconstruction a set of basic criteria was developed.
The aim of these requirements was to select positively charged particles 
with a precise measurement of the absolute value of the particle rigidity and velocity.
Furthermore, events with more than one track, likely to be products of hadronic interactions
occurring in the top part of the apparatus, were rejected. 

{%\color{red}
Events were selected requiring:
\begin{itemize}
\item A single track fitted within the spectrometer fiducial volume
where the reconstructed track is at least 1.5 mm away from the magnet walls.
\item A positive value for the reconstructed track curvature.
\item Selected tracks must have at least 
4 hits on the $x$-view and at least 3 hits on the $y$-view to ensure a good rigidity reconstruction.
\item A maximum of one hit paddle in the two top planes of the ToF system. 
\item The hit paddles in S1 and S2 must match the extrapolated trajectory from the spectrometer. 
\item A positive value for the measured time of flight. This selection ensures that the particle enters \pam\ from above.
\item For the selection of the hydrogen sample
no activity in the CARD and CAT scintillators of the anticoincidence system is required.
\end{itemize}
The anticoincidence selections on the hydrogen sample were necessary} since most secondary
particles that entered the \pam\ fiducial acceptance were by-products of hadronic interactions taking place
in the aluminium dome or in the S1 and S2 scintillators. Such particles were generally accompanied by other secondary particles
which hit the anticoincidence detectors. For the selection of the helium sample there were no anticoincidence 
requirements since contamination by secondary helium coming from heavier nuclei spallation is negligible.

\subsubsection{Galactic particle selection} \label{sec:galactic}
The Resurs-DK1 satellite orbital information was used to estimate the local geomagnetic cutoff, $G$, in
the St\"{o}rmer approximation~\citep{shea} using the IGRF magnetic field model \citep{2005EP&S...57.1135M} along the orbit.
%To select the primary (galactic) cosmic ray component the local geomagnetic cutoff, $G$, in
%the St\"{o}rmer approximation~\citep{shea} was 
%evaluated. The  value of $G=14.9/L^2$, valid for vertically incident particles, is 
%estimated using the IGRF magnetic field model along the orbit.
%From this the McIlwain $L$ shell is calculated~\citep{2005EP&S...57.1135M}. 
The maximum zenith angle for events entering the \pam\ acceptance
was 24 degrees with a mean value of 10 degrees.
To select the primary (galactic) cosmic ray component particles were binned by requiring that $\rho_m > k \cdot G$,
%their energy at the top of the payload (ToP) be greater than $k$ times the geomagnetic cutoff, 
where $\rho_m$ is the lowest edge of the rigidity interval and $k=1.3$ is a safety factor required 
to remove any directionality effects due to the Earth's penumbral regions.
{%\color{red}
Galactic particles losing energy while crossing the detector may be rejected by this selection. 
This effect is accounted for using Monte Carlo simulations. 
}

\subsubsection{Charge selection} \label{sec:chargesel}
\begin{figure}[t]
    \centering
    \plotone{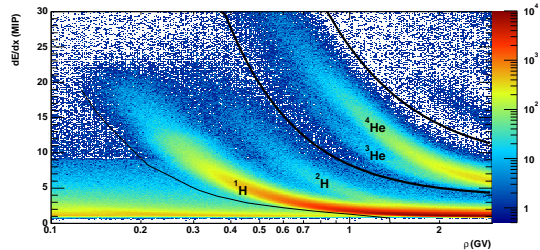}
    \caption{Energy loss in the silicon detectors of the tracking system (mean energy deposition in all planes hit) as a function of reconstructed rigidity for positively charged
      particles. The hydrogen and helium bands are clearly visible. The black lines represent
      the selection for H and He nuclei.}
    \label{im:dedx}
\end{figure}

{%\color{red}
Particle charge identification relies on the ionization measurements provided by the
magnetic spectrometer. Depending on the number of hit planes there can be up to 12
$dE/dx$ measurements. The arithmetic mean of those measurements is shown in Fig. \ref{im:dedx}. 
A rigidity dependent selection on the mean $dE/dx$ from the spectrometer is used
to select $Z=1$ or $Z=2$ candidates and is depicted by the solid lines in Fig. \ref{im:dedx}.

%The effect of misidentified \hef\ wrongly reconstructed as $Z=1$ particles could result in
%a significant contamination to the \deu\ sample.
%Misidentified helium was measured performing the charge identification
%using strict selections on the $dE/dx$ measurement of the six ToF planes 
%and then applying the $Z=1$ request in the spectrometer.
The residual contamination of $Z=2$ particles in the $Z=1$ sample was studied
selecting helium events using the $dE/dx$ information from the ToF system, 
and then applying the $Z=1$ selection previously described.
The fraction of misidentified helium events was found to be less than $10^{-4}$.
Since  the \deu/\hef\ ratio is roughly 0.15 the resulting contamination in the \deu\ sample
from misidentified \hef\ was estimated less than $10^{-3}$.
Similarly the $Z=1$ residual contamination  as well as contamination by heavier nuclei 
in the $Z=2$ sample was estimated to be negligible.
}
%The selectino bands are optimized in order to have the highest efficiency while mantaining no evidence of charge misidentification.

\subsection{Isotope separation}
%The separation between the different isotopes with mass $m$ and charge $Z$ at fixed rigidity is guaranteed 
%by
%\begin{equation}
% \frac{1}{\beta} = \sqrt{1+\frac{m^2}{Z^2 \rho^2}}
%\label{eq:beta_r}
%\end{equation} 
%and decreases with increasing energy. The ToF resolution allows \pam\ to resolve \deu\ and \het\ up
%to rigidities $\sim 2.5$ GV as shown in Fig. \ref{im:beta_r}
\begin{figure}[t]
    \centering
    \plotone{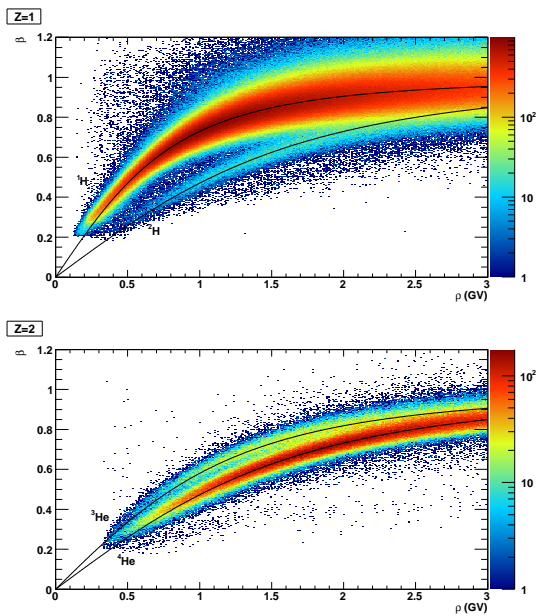}
    \caption{Mass separation for $Z=1$ (\emph{top}) and $Z=2$ (\emph{bottom}) particles using the $\beta$-rigidity method. 
      The black lines represent the expectations for each isotope.}
    \label{im:beta_r}
\end{figure}

\begin{figure}[t]
    \centering
    %\epsscale{0.9}
    \plotone{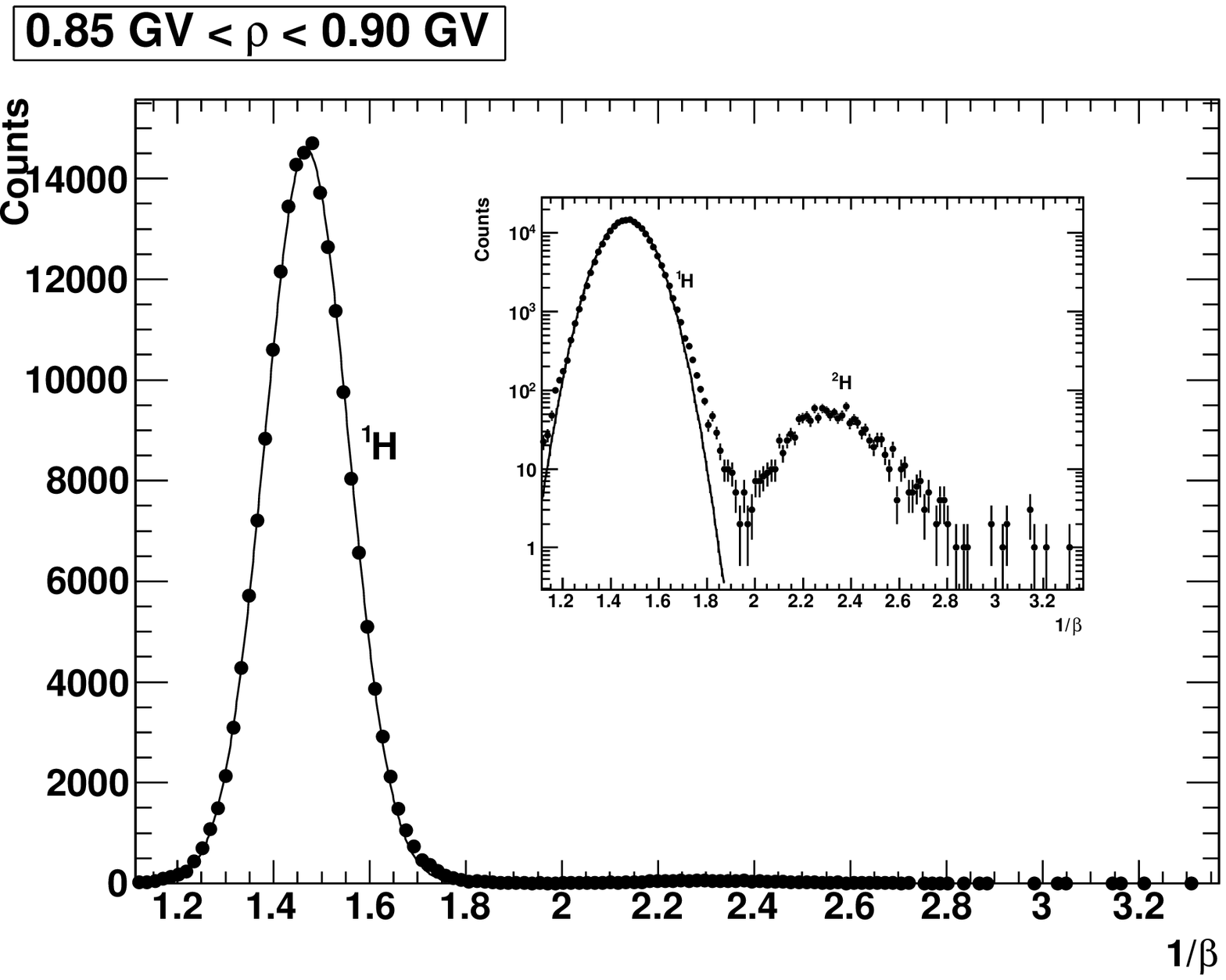}
    \plotone{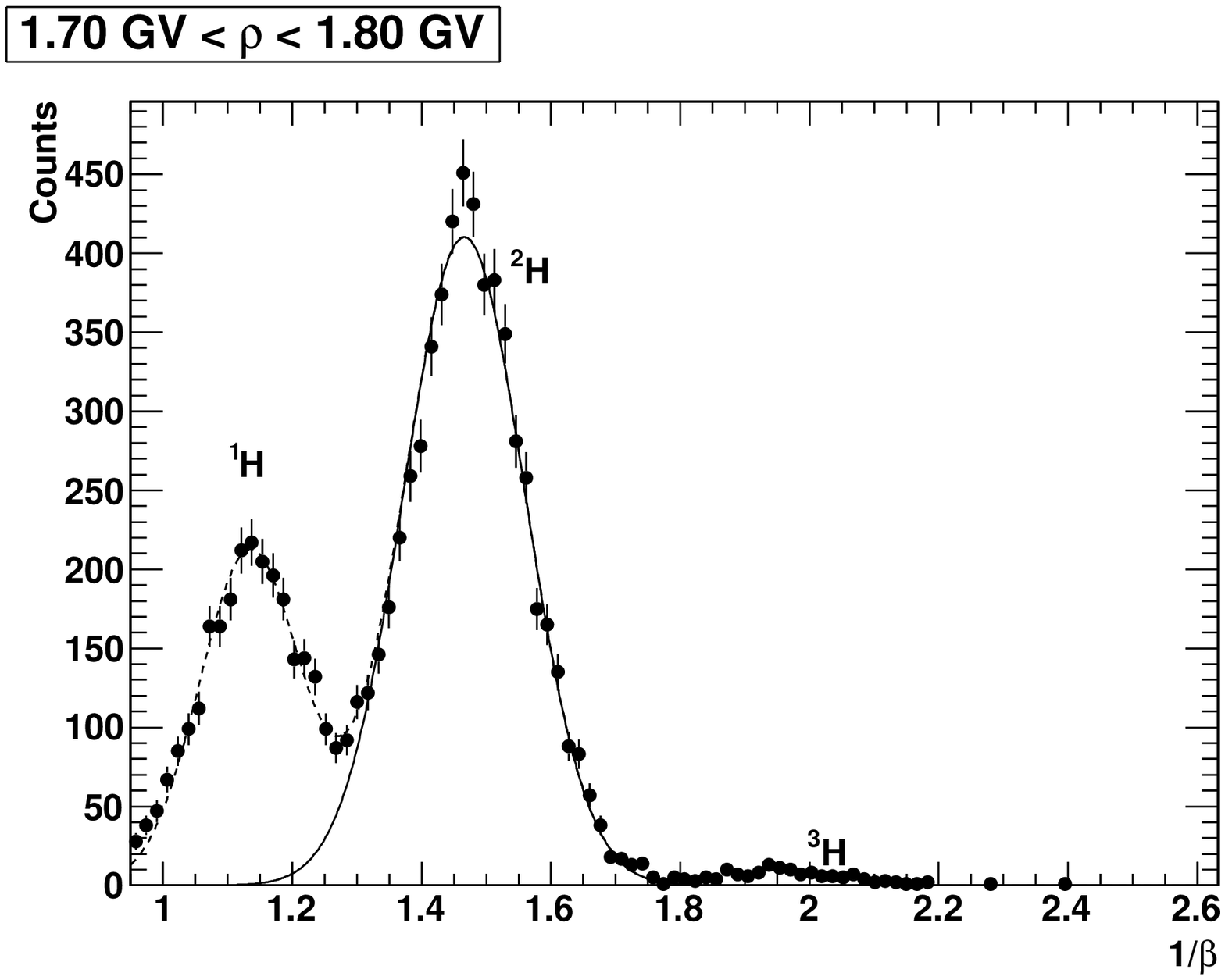}
    \caption{Examples of $1/\beta$ distributions for hydrogen in the 0.329 - 0.361 GeV/n
    kinetic energy range for \prot\ (\emph{top})
      and \deu\ (\emph{bottom}). The solid line shows the estimated \prot\ and \deu\ signal while the
      dashed line shows the combined fit used in the \deu\ case to improve the fit result. 
      The box in the upper plot shows the same distribution with a logarithmic scale to show the \deu\ component that
      is not visible on a linear scale.
	  The \prot\ component in the \deu\ distribution in the bottom plot is suppressed by the additional selection shown in Fig \ref{im:trk_lowest}.      
      The hydrogen sample also contains a small
      fraction of \textsuperscript{3}H events which are used to test the reliability of the Monte Carlo simulation. }
    \label{im:beta_fit_h}
\end{figure}

\begin{figure}[t]
    \centering
    %\epsscale{0.9}
    \plotone{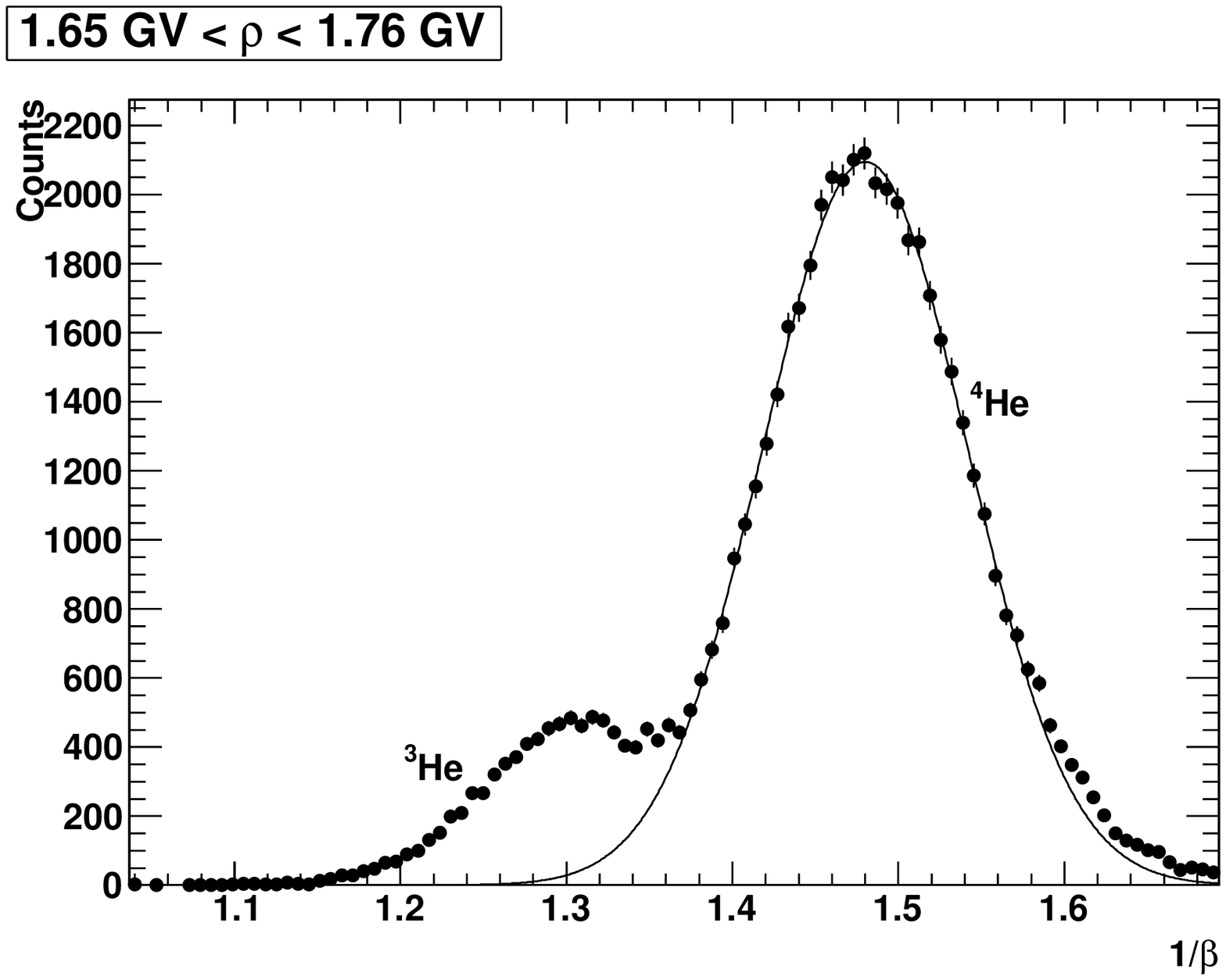}
    \plotone{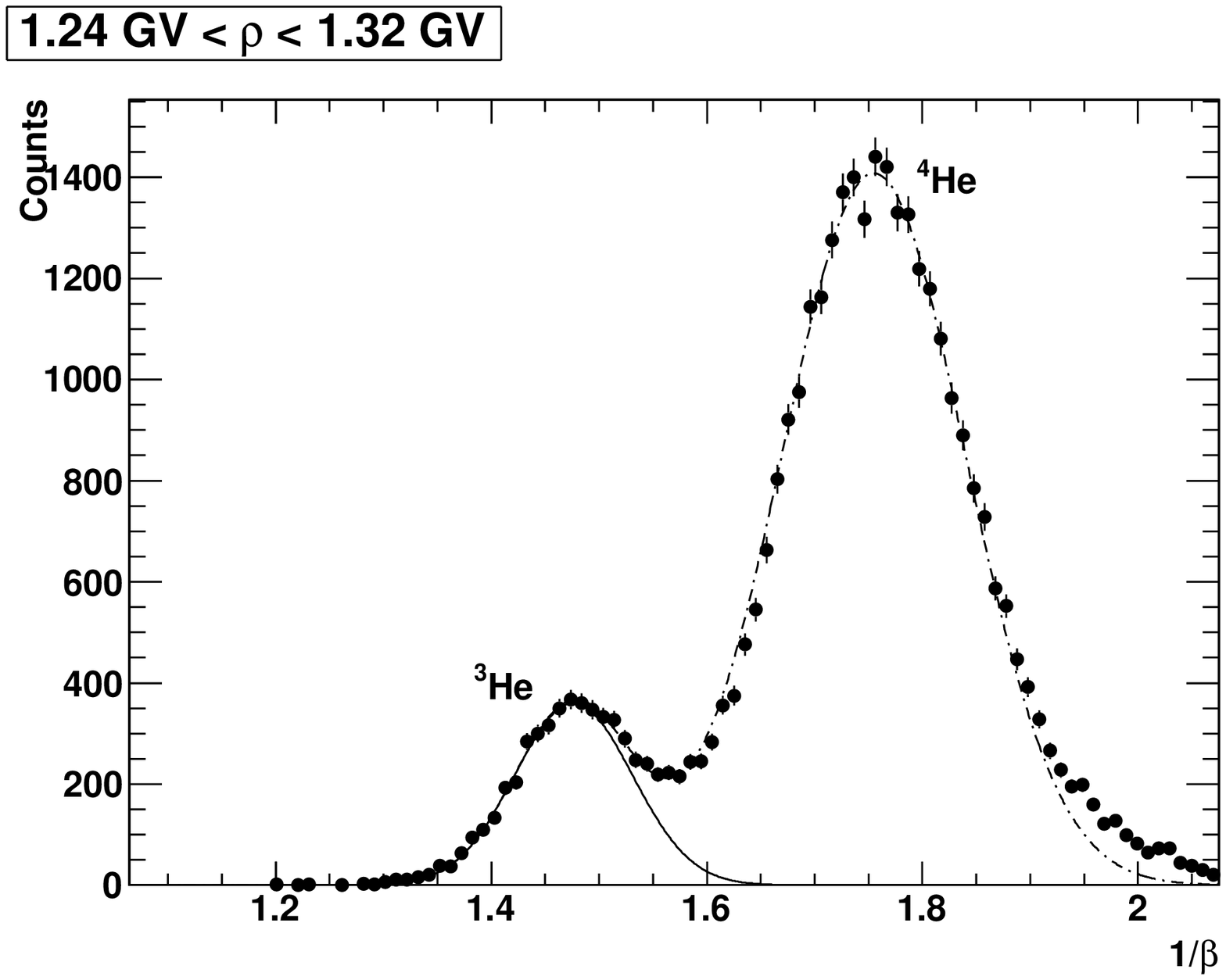}
    \caption{Examples of $1/\beta$ distributions for helium in the 0.312 - 0.350 GeV/n kinetic
    energy range for \hef\ (\emph{top})
      and \het\ (\emph{bottom}). The solid line shows the estimated \hef\ and \het\ signal while the
      dashed line shows the combined fit used in the \het\ case to improve the fit result.
	 }
    \label{im:beta_fit_he}
\end{figure}

\begin{figure}[t]
    \centering
    \plotone{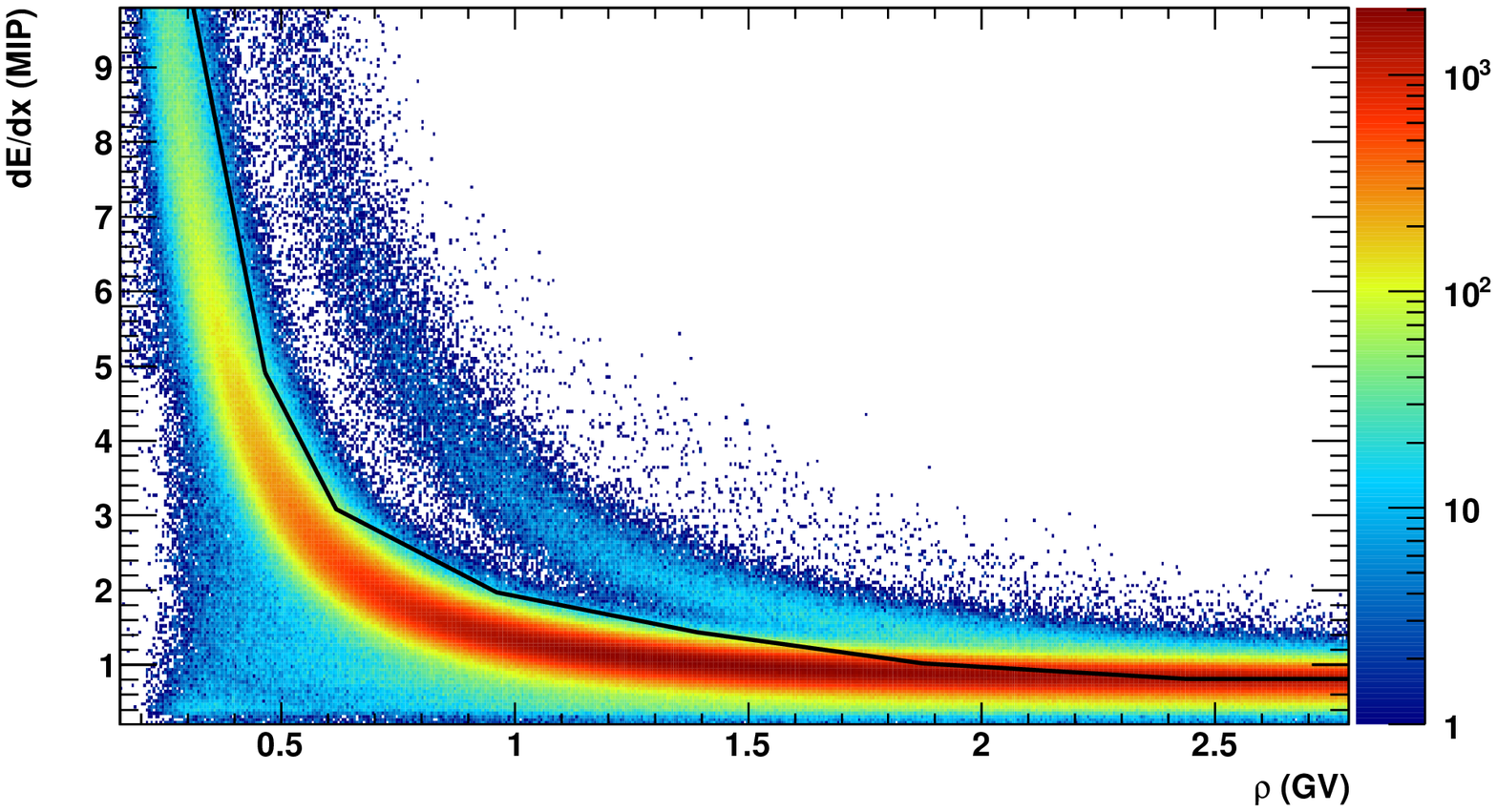}
    \caption{The lowest of the 12 energy releases in the tracking system as a function of rigidity for $Z=1$ events. Events with energy releases above the black line were selected
    for the \deu\ analysis.}
    \label{im:trk_lowest}
\end{figure}

{%\color{red}
The selection criteria described in the previous section provided
clean samples of $Z = 1$ and $Z=2$ particles. 
{%\color{red}
%Velocity information was used to separate isotopes. 
In each sample an isotopic separation at fixed 
rigidity is possible by reconstructing $\beta$, where
\begin{equation}
  \beta = \left( 1 + \frac{m^2}{Z^2 \rho^2} \right)^{-1/2}
\end{equation}
as shown in
Fig.~\ref{im:beta_r} for events in the
($\beta$, $\rho$) plane.
}
%shows the measured $\beta$ for the selected $Z = 1$ and 2 particles as
%a function of rigidity. Isotopes are resolved up to $\sim$ 2
%GV.

Isotope separation as well as the determination of isotope fluxes was
performed in intervals of kinetic energy per nucleon. Since the
magnetic spectrometer 
measures the rigidity of particles, this implies different rigidity
intervals according to the isotope under study. For example
Fig.~\ref{im:beta_fit_h} shows the 1/$\beta$ distributions used to select \prot\ 
(top panel) and \deu\ (bottom panel) in the kinetic energy interval
0.329 - 0.361 GeV/n corresponding to 0.85 - 0.9 GV for \prot\ and 1.7 - 1.8
GV for \deu. 
Particle counts were subsequently extracted from a Gaussian fit to the
1/$\beta$ 
distribution in each rigidity range as shown by the solid lines in
Fig.~\ref{im:beta_fit_h}. 

Separation between \het\ and \hef\ was obtained in a similar way.
Fig.~\ref{im:beta_fit_he} shows the 1/$\beta$ distributions 
used to select \het\
(bottom panel) and \hef\ (top panel) in the kinetic energy interval
0.312 - 0.350 GeV/n corresponding to 1.24 - 1.32 GV for \het\ and 1.65 - 1.76
GV for \hef.  

It should be noted that, because of the large proton background, an
additional selection, based 
on the lowest energy 
release among the 12 measurements provided by the tracking system
(often referred as {\em truncated mean}),
was used to produce the 1/$\beta$ distributions in the \deu\ case.
Fig.~\ref{im:trk_lowest} shows this quantity for $Z = 1$ particles. The solid line
indicates the condition on the minimum energy release used for the
selection. 

The selected number of hydrogen and helium events are summarized in the 
second and third column of Tables \ref{table:events_h1} and \ref{table:events_he}.
}

{%\color{red} Insert event table here!!
\input{table1}
\input{table2}
}

\subsection{Flux determination}

The procedure described in the previous section was used to estimate the number of \prot\ and \deu\ events in
the $Z=1$ sample and the number of \het\ and \hef\ events in the $Z=2$ sample. 
%In order to derive the corresponding
%flux for each isotope corrections had to be performed.
%Several of them were obtained using 
%Isotope fluxes were corrected using
To derive each isotope flux the number of selected events
had to be corrected for the selections efficiencies, particle losses, 
contamination and energy losses. These corrections were 
obtained using
a Monte Carlo simulation of the \pam\ apparatus based on the 
\texttt{GEANT4} code \citep{Geant4} and from the flight data.
{%\color{red}
The simulation contains an accurate representation of the geometry and performance of the \pam\ detectors. 
The measured noise of each silicon plane of the spectrometer and its performance variations over the duration 
of the measurement were accounted for. The simulation code was validated by comparing the distributions of 
several significant variables with those obtained from real data.
}
Hadronic interactions for all the isotopes under study were handled via the \texttt{QGSP\_BIC\_HP} physics list.
\input{table3}

\begin{figure}[t]
    \centering
    \plotone{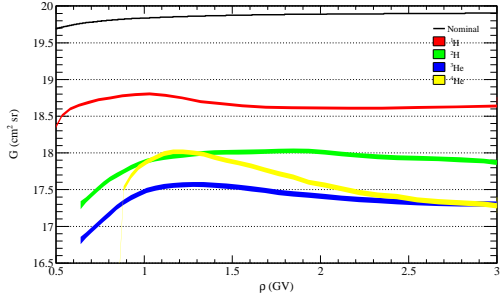}
    \caption{The nominal geometrical factor $G_F$ as a function of rigidity (solid line). The filled bands represent the effective geometrical factor $G(E)$ for each isotope
    with the associated uncertainty.}
    \label{im:geofa}
\end{figure}

\begin{figure}[t]
    \centering
    \plotone{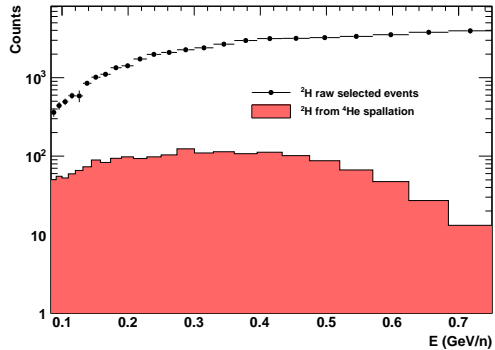}
    \caption{Raw counts for \deu\ collected by \pam\ in 2006 together with the expected 
      contamination from \hef\ spallation in the top part of the apparatus.}
    \label{im:contamination}
\end{figure}

The following corrections to the number of selected events were applied:
\begin{itemize}
\item {\em Selection efficiencies:} The redundant information provided by \pam\ allowed most of the 
selection efficiencies to be estimated directly from flight data. 
For example the efficiency of the charge selections was evaluated on a sample
of events selected with the ToF $dE/dx$ measurements in the same
way as described in section \ref{sec:chargesel} for the charge misidentification study.
The efficiency of the tracking system
was however obtained from the Monte Carlo simulation. 
%The overall efficiency of all selections during the second half of 2006 
%is $\simeq78\%$ for \prot, $\simeq73\%$ for \deu, $\simeq75\%$ for \het, and $\simeq79\%$ for \hef;
%while during 2007 the overall efficiency 
%is $\simeq56\%$ for \prot, $\simeq51\%$ for \deu, $\simeq52\%$ for \het, and $\simeq56\%$ for \hef, the details are summarized in Table~\ref{table:eff}.
The decrease in efficiency from 2006 to 2007 is due to the failure of some of the front-end chips in the tracking system.
This situation was included in the Monte Carlo simulation, as discussed in \cite{2013ApJ...765...91A}.
The efficiencies of the various selections are reported in Table \ref{table:eff}.

\item {\em Hadronic interactions:} Helium and hydrogen nuclei may be lost due to hadronic interactions in the 2~mm thick aluminium pressurized 
container and the top scintillator detectors. The correction to the flux due to this effect was included in the
\pam\ geometrical factor as follows:
\begin{equation}
  G(E) = \left[1 - b(E)\right] G_F
\end{equation} 
where $G(E)$ is the effective geometrical factor used for the flux determination, 
$b(E)$ is a correction factor which accounts for the effect of inelastic scattering, 
$G_F$ is the nominal geometrical factor which is almost constant above 1 GV and slowly decreases by $\sim2\%$ to lower energies, 
where the particle trajectory in the magnetic field is no longer straight.
The requirement on the fiducial volume corresponds to a geometrical factor $G_F = 19.9$  cm\textsuperscript{2} sr above 1 GV. 
The correction factor $b(E)$ is different for each isotope and has been derived from the Monte Carlo simulation, being
$\simeq 6\%$ for protons, $\simeq 10\%$ for deuterium, and $\simeq 13\%$ for both helium isotopes.
The nominal geometrical factor and the effective geometrical factor for each isotope are shown in Fig. \ref{im:geofa}.

\item {\em Contamination:} The 
contribution to \deu\ from \hef\ inelastic scattering was evaluated from the simulation 
(Fig. \ref{im:contamination}) and subtracted from the raw \deu\ counts (see column 4 of Table \ref{table:events_h1}).
The contamination in the \het\ sample from \hef\ fragmentation was also 
evaluated and it was estimated to be less than $1\%$. This was included in the 
systematic uncertainty of the measurement. 

\item {\em Energy loss and resolution:} The finite resolution of the magnetic spectrometer and particle slowdown 
due to ionization energy losses results in a distortion of the particle spectra. 
A Bayesian unfolding procedure, described
in \cite{dagostini}, was used to derive the number of events at the top of the payload (see \cite{2011Sci...332...69A_red}).
\end{itemize}

The flux was then calculated as follows:
\begin{equation}
%  \Phi_{\text{ToP}} (E) = \frac{1}{\varepsilon (E)}\frac{N_\text{ToP}(E)}{T G(E) \Delta E}
  \Phi_{\text{ToP}} (E) = \frac{N_\text{ToP}(E)}{T G(E) \Delta E}
\end{equation}
where $N_\text{ToP}(E)$ is the unfolded particle count for energy $E$, also corrected for all the selection efficiencies
(see Tables \ref{table:events_h1} and \ref{table:events_he}, rightmost two columns), 
%$\varepsilon(E)$ is the overall selection efficiency, 
$\Delta E$ is the energy bin width, and $G(E)$ is the effective geometrical factor. 
The live time, $T$, as evaluated by the trigger system, depends on the orbital selection as described in section \ref{sec:galactic}
(e.g. see \cite{bru08}). The live and dead time are cross-checked with the 
total acquisition time measured by the on-board CPU to remove possible systematic effects
in the time counting.

\subsection{Systematic uncertainties}
{%\color{red}
The possible sources of systematic uncertainties considered in this analysis are listed below and are also included 
in Tables \ref{tab:hydrogen} and \ref{tab:helium} and in Figs. \ref{im:fig_data_h}, \ref{im:fig_data_he} and \ref{im:ratios}.
\begin{itemize}
  \item {\em Quality of the $1/\beta$ fit:} The quality of the Gaussian fit procedure was tested
  using the truncated mean of the energy deposited 
  in the electromagnetic calorimeter to select pure samples of \prot\ and \deu\ from non-interacting events. 
  The two samples were then merged to form a control sample
  for the fitting algorithm. The number of reconstructed events from the Gaussian fit was found to agree 
  with the number of events selected with the calorimeter, so no systematic uncertainty was assigned
  to this procedure. 

  \item {\em Selection efficiencies:} The estimation of the selection efficiencies is affected by a statistical error due to 
  the finite size of the sample used for the efficiency evaluation. This error was considered and 
  propagated as a systematic uncertainty. For the efficiency of the ToF and AC selections this uncertainty is
  $0.21\%$ at low energy (120 MeV/n) and drops to $0.14\%$ at high energy (600-900 MeV/n).
  For the tracker selections the uncertainty is $0.3\%$ at low energy increasing to $0.4\%$ at high energy.

  \item {\em Galactic particle selection:} The correction for particles lost due to this selection has an
   uncertainty, due to the size of the Monte Carlo sample, 
   which decreases from $6\%$ to $0.07\%$ as the energy increases from 120 MeV/n to 900 MeV/n.

  \item {\em Contamination subtraction:} The subtraction of the contamination 
  results in a systematic uncertainty on the \deu\ flux of $1.9\%$ at low energy
  dropping below $0.1\%$ at 300 MeV/n due to the finite size of the Monte Carlo sample.
  To test the validity of the Monte Carlo simulation the \textsuperscript{3}H component, identified as
  the additional cluster of events at low $\beta$ in the hydrogen sample visible in Fig. \ref{im:beta_fit_h}, 
  was used. The \textsuperscript{3}H events are created by \hef\ spallation in the 
  top part of the apparatus since no tritium of galactic origin should survive propagation to Earth.
  The observed number of \textsuperscript{3}H events was used to test that the Monte Carlo simulation 
  correctly inferred the number of \deu\ events coming from \hef\ fragmentation. 
  For example, for the 2006 data-set in the rigidity range
  between $1.7$ GV and $1.8$ GV the flight data sample contains $136 \pm 17$ tritium events,
  while $110 \pm 15$ events are expected according to the Monte Carlo simulation (Fig. \ref{im:beta_fit_h}). 
  Simulation and flight data were in agreement within a 10\% tolerance.
  This discrepancy was treated as an
  additional systematic uncertainty on the estimated number of contamination events. 
  The 10\% systematic uncertainty on the \deu\ contamination translates in an additional 1\% uncertainty
  on the number of reconstructed \deu\ events.
    
  \item {\em Geometrical factor:} The uncertainty on the effective geometrical factor 
%  due to the size of the simulation sample used, 
  as estimated from the Monte Carlo simulation is almost independent of energy and amounts to $0.18\%$.
  
  \item {\em Unfolding procedure:} 
%  The systematic uncertainty on the unfolding procedure was evaluated
%  generating events with the Monte Carlo simulation according to a known energy spectrum. The sample of
%  events obtained was then split into two independent halves. One has been used to derive the smearing matrix
%  (see \cite{dagostini} for all the details) and the other half has been used to construct the event sample to unfold.
%  The difference between the unfolded MC spectrum and the true spectrum used to generate the events has been
%  taken as a systematic uncertainty and amounts to $2\%$ constant in energy.
  As discussed in \cite{2011Sci...332...69A_red} two possible systematic effects have been studied regarding the unfolding procedure: 
  the uncertainty associated with the simulated smearing matrix and the intrinsic accuracy of the procedure. 
  The former was constrained by the checking for compatibility between measured and simulated spatial residuals 
  and was found to be negligible. The latter was estimated by folding and unfolding a known spectral shape 
  with the spectrometer response and was found to be 2\%, independent of energy.
\end{itemize}

}

%This allows to estimate a systematic
%uncertainty for the fit procedure which takes also into account for the
%non-Gaussian tails that can be seen in the $1/\beta$ distributions of Fig.~\ref{im:beta_fit_h} and \ref{im:beta_fit_he}.
	
%Spostare in 3.4 Systematics!

\section{Results}
%The asbolute fluxes of \prot, \deu, \het\ and \hef\ measured by the \pam\ experiment from July 2006 to
%December 2007 are presented in Fig. \ref{im:fluxes} while the three isotopic ratios \deu/\prot, 
%\het/\hef\ and \deu/\hef\ are shown in Fig. \ref{im:ratios} together with previous measures
%\citep{2002ApJ...564..244W, 1998ApJ...496..490R, 2011ApJ...736..105A} and with theoretical predictions
%based on the numerical solution of the Parker transport equation \citep{galprop54, 2006ApJ...642..902P}.
%We compared the experimental results with two models already presented in literature: the solid lines in
%the plots refer to a cosmic ray propagation model in presence of a second order reacceleration, while
%all dashed lines refer to a model in which only the diffusion process occur during propagation.
%A detailed description of the two models can be found in the reference paper.

\begin{figure*}[t]
    \centering
    \epsscale{0.75}
    \plotone{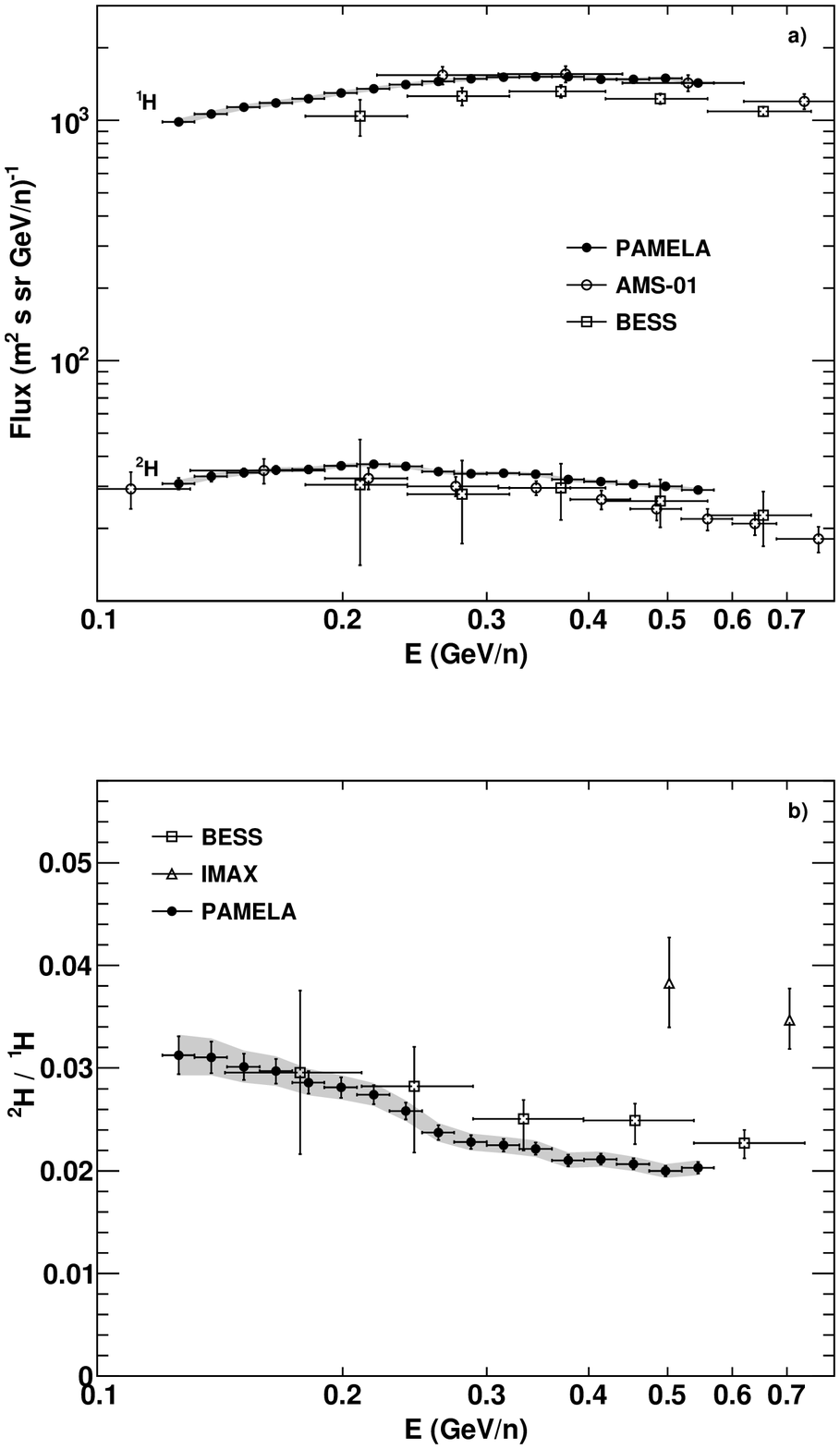}
    \caption{\prot\ and \deu\ absolute fluxes (a) and their ratio (b) compared to previous experiments:  AMS-01 \citep{2002PhR...366..331A,2011ApJ...736..105A_red}, BESS \citep{2002ApJ...564..244W}, IMAX \citep{1998ApJ...496..490R_red}. 
      Error bars show the statistical uncertainty while shaded areas show the systematic uncertainty. }
    \label{im:fig_data_h}
\end{figure*}

\begin{figure*}[t]
    \centering
    \epsscale{0.75}
    \plotone{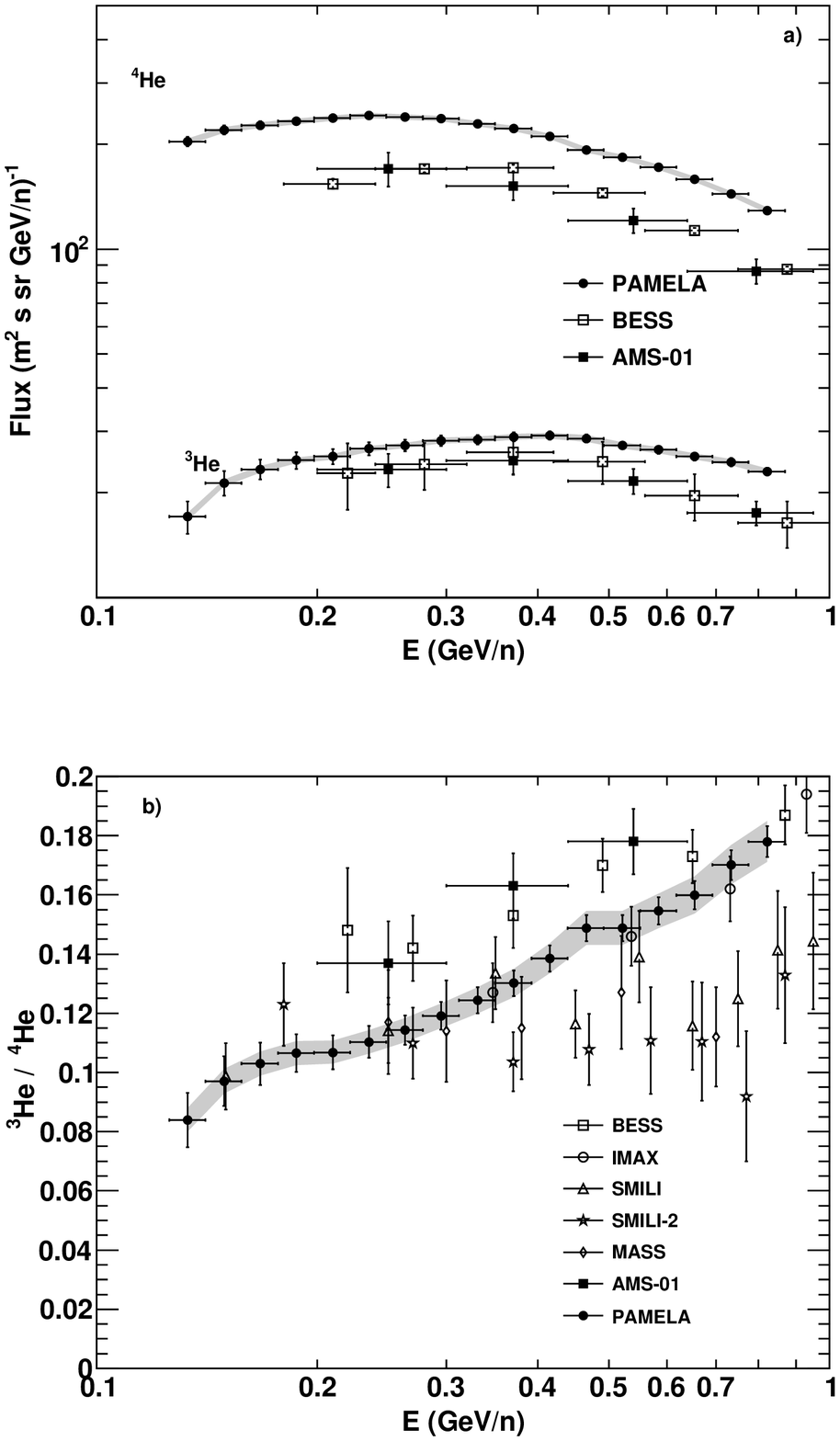}
    \caption{\hef\ and \het\ absolute fluxes (a) and their ratio (b) compared to previous experiments: AMS \citep{2011ApJ...736..105A_red}, BESS \citep{2002ApJ...564..244W}, IMAX \citep{1998ApJ...496..490R_red}, SMILI-2\citep{1995ICRC....2..630W}, MASS \citep{1991ApJ...380..230W}, SMILI-1 \citep{1993ApJ...413..268B_red}. Error bars show statistical uncertainty while shaded areas show systematic uncertainty.}
    \label{im:fig_data_he}
\end{figure*}

\begin{figure*}[t]
    \centering
    \epsscale{0.75}
    \plotone{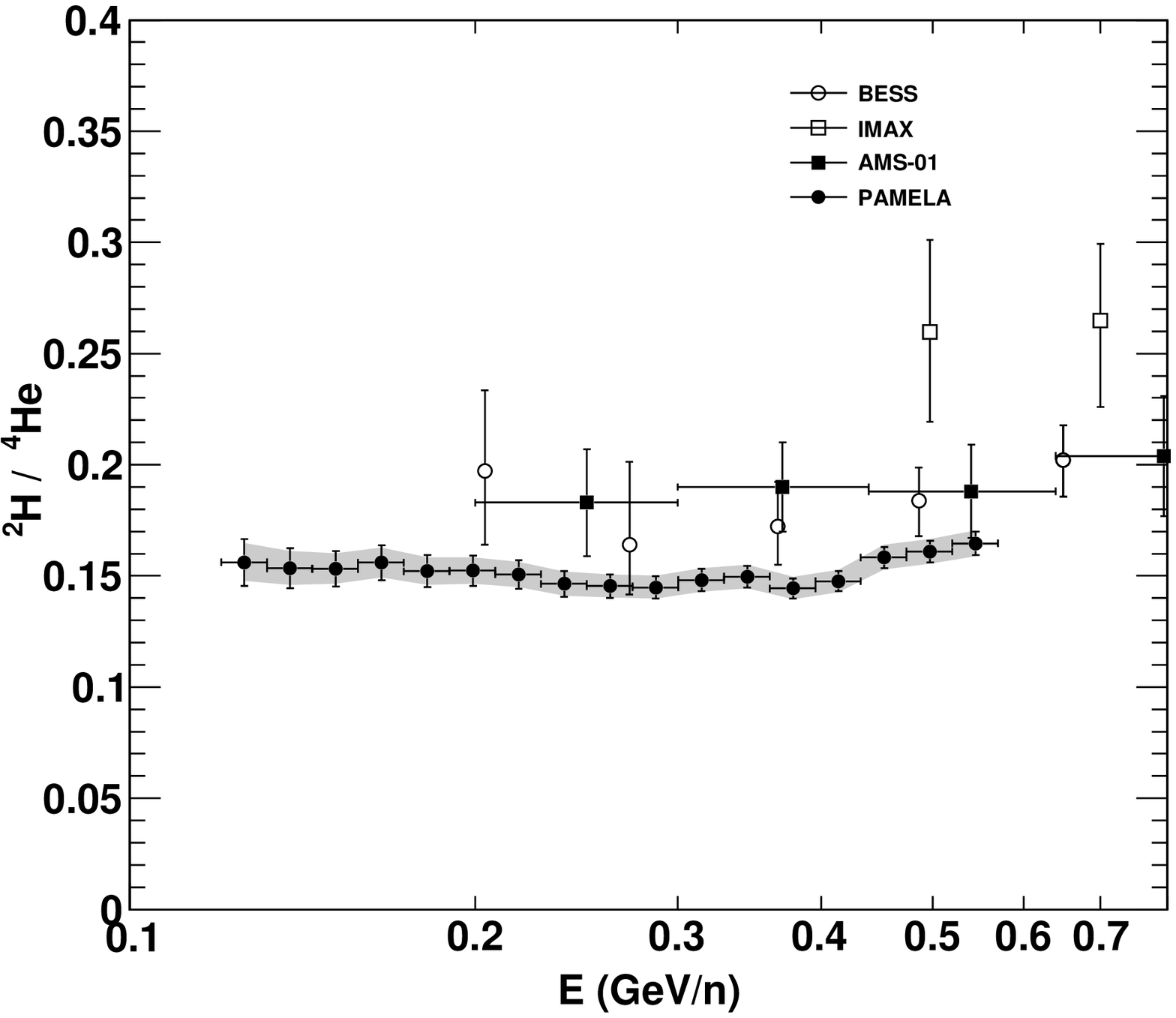}
    \caption{\deu/\hef\ ratio compared to previous experiments:  AMS-01 \citep{2011ApJ...736..105A_red}, BESS \citep{2002ApJ...564..244W}, IMAX \citep{1998ApJ...496..490R_red}. 
      Error bars show statistical uncertainty while shaded areas show systematic uncertainty. }
    \label{im:ratios}
\end{figure*}

Figure~\ref{im:fig_data_h} and \ref{im:fig_data_he} show hydrogen and helium isotope
fluxes (top) and the ratios of the fluxes (bottom). Results are also reported in Tables 
\ref{tab:hydrogen} and \ref{tab:helium}. Fig. \ref{im:ratios} shows the \deu/\hef\ ratio as a function of kinetic energy per nucleon, compared to previous measurements~\citep{2011ApJ...736..105A_red,2002ApJ...564..244W,1998ApJ...496..490R_red,1995ICRC....2..630W,1991ApJ...380..230W,1993ApJ...413..268B_red}. 
%and theoretical predictions based on a
%calculation using GALPROP \citep{webgalprop_red} with 
%propagation parameters from \cite{2006ApJ...642..902P_red}. 
%{%\color{blue} 
%The two models implemented in GALPROP are chosen since they give an overall good fit
%of previous cosmic ray data \citep{2006ApJ...642..902P_red}: the B/C ratio and the proton (at least up to a few hundred GeV), 
%positron, and antiproton energy spectra.
%The calculated fluxes were modulated using the force-field approximation \citep{1968ApJ...154.1011G}
%with a solar modulation parameter
%$\Phi = 410$~MV for the DR model and $\Phi = 543$~MV for the PD model 
%obtained from a comparison of the calculated and measured proton spectra. 
%Two different solar modulation parameters are used in the two models
%since the $\Phi$ parameter 
%is not really a physical quantity that can 
%in this approximation cannot
%be estimated from the knowledge of other heliospheric observables
%but it is rather a proxy of the changes in the solar activity over time.
%Both models are able to describe the \prot\ spectrum obtained in this work.  
It is worth noting that flux ratios (in particular
the \het/\hef\ ratio), if modulated using the force-field approximation \citep{1968ApJ...154.1011G}, show very little dependence on solar activity and can therefore 
be used to  
discriminate between various propagation models of GCR in the Galaxy.
%}
%The larger discrepancy between calculated and measured
%spectra for helium than for hydrogen data is because the 
%hydrogen and helium spectral shapes are identical in the calculation. As discussed
%in \cite{2011Sci...332...69A_red}, \pam\ data show clear differences in
%the spectral shape of the two species.

The \pam\ results are the most precise to date.
Considering the relatively large spread in the existing data, \pam\
results agree with previous 
measurements, in particular with BESS results
for $^2$H and IMAX results for $^3$He.
Previous measurements are affected
by large uncertainties and, for $^3$He where more measurements are
available, there is a large spread between data.   
All the measurements displayed in Figures~\ref{im:fig_data_h},
\ref{im:fig_data_he}, and \ref{im:ratios}, except 
AMS-01, are from balloon-borne experiments and are affected by a
non-negligible background of atmospheric secondary particles. 

A high precision measurement of the H and He isotope quartet
abundances represents a significant step forward in
modelling the origin and propagation of GCRs. 
The constraints on diffusion-model parameters set by the quartet ($^1$H, $^2$H, $^3$He,
and $^4$He) were recently revisited \citep{2012A&A...539A..88C}.  
It was found that the constraints on the parameters were competitive with 
those obtained from the 
B/C flux ratio analysis and available data supported 
the universality of GCR propagation in the Galaxy.  
The tightest constraint was obtained when the He flux was included in
the fit. This is because at energies of a
few GeV about 10$\%$ of He
is from fragmentation of heavier nuclei, which is a non-negligible amount
given the precision (1$\%$ statistical) of the \pam\ He data \citep{2011Sci...332...69A_red}.  
%The constraint on propagation parameters is still limited by the large
%discrepancy among different $^2$H and $^3$He and improvements are
%expected if more precise data will be available. 
\input{table4.tex}

\input{table5.tex}

\end{document}

%% file: table1.tex
\begin{deluxetable*}{cccccc}
\tablecaption{Number of selected and reconstructed events for the $Z=1$ sample.\label{table:events_h1}}
\tabletypesize{\scriptsize} 
\tablewidth{0pt}
%\hline
%\hline
\tablehead{ \multicolumn{1}{c}{Kinetic energy} & \multicolumn{1}{c}{\prot\ events} & \multicolumn{1}{c}{\deu\ events} &   \multicolumn{1}{c}{\deu\ contamination} &
                     \multicolumn{1}{c}{\prot\ extrapolated events} & \multicolumn{1}{c}{\deu\ extrapolated events} \\
\multicolumn{1}{c}{per nucleon (GeV/n)} & \multicolumn{3}{c}{} &  \multicolumn{1}{c}{at top of payload ($N_\text{ToP}$)} &  \multicolumn{1}{c}{at top of payload ($N_\text{ToP}$)} }
%\hline
\startdata
$0.120 - 0.132 $ &  $38927 \pm 477$ & $2179 \pm 99$ & $207 \pm 9$ & $70273 \pm 867$ & $3474 \pm 187$\\
$0.132 - 0.144 $ &  $47689 \pm 521$ & $2401 \pm 60$ & $212 \pm 5$ & $85115 \pm 940$ & $4120 \pm 106$\\
$0.144 - 0.158 $ &  $58104 \pm 587$ & $2815 \pm 114$ & $230 \pm 9$ & $102825 \pm 1049$ & $4800 \pm 209$\\
$0.158 - 0.173 $ &  $70390 \pm 635$ & $3284 \pm 120$ & $248 \pm 9$ & $122634 \pm 1115$ & $5594 \pm 225$\\
$0.173 - 0.190 $ &  $84622 \pm 703$ & $3752 \pm 124$ & $261 \pm 9$ & $145937 \pm 1227$ & $6397 \pm 227$\\
$0.190 - 0.208 $ &  $102408 \pm 781$ & $4312 \pm 134$ & $276 \pm 9$ & $177148 \pm 1369$ & $7385 \pm 246$\\
$0.208 - 0.228 $ &  $123869 \pm 843$ & $4865 \pm 144$ & $286 \pm 8$ & $212463 \pm 1465$ & $8343 \pm 265$\\
$0.228 - 0.250 $ &  $145923 \pm 929$ & $5505 \pm 153$ & $298 \pm 8$ & $248196 \pm 1601$ & $9298 \pm 278$\\
$0.250 - 0.274 $ &  $173016 \pm 996$ & $6015 \pm 159$ & $302 \pm 8$ & $291879 \pm 1695$ & $10165 \pm 287$\\
$0.274 - 0.300 $ &  $200767 \pm 1105$ & $6683 \pm 173$ & $312 \pm 8$ & $336058 \pm 1868$ & $11269 \pm 310$\\
$0.300 - 0.329 $ &  $231937 \pm 1129$ & $7419 \pm 192$ & $324 \pm 8$ & $388972 \pm 1903$ & $12635 \pm 345$\\
$0.329 - 0.361 $ &  $266950 \pm 1259$ & $8274 \pm 209$ & $339 \pm 9$ & $443768 \pm 2107$ & $14143 \pm 374$\\
$0.361 - 0.395 $ &  $303941 \pm 1354$ & $9356 \pm 227$ & $358 \pm 9$ & $506860 \pm 2291$ & $15447 \pm 402$\\
$0.395 - 0.433 $ &  $343790 \pm 1389$ & $10153 \pm 245$ & $358 \pm 9$ & $568056 \pm 2315$ & $17005 \pm 432$\\
$0.433 - 0.475 $ &  $382305 \pm 1511$ & $10922 \pm 259$ & $345 \pm 8$ & $630535 \pm 2517$ & $18588 \pm 461$\\
$0.475 - 0.520 $ &  $424071 \pm 1583$ & $11551 \pm 287$ & $310 \pm 8$ & $703994 \pm 2654$ & $20277 \pm 511$\\
$0.520 - 0.570 $ &  $466939 \pm 1658$ & $12425 \pm 337$ & $258 \pm 7$ & $775857 \pm 2784$ & $22210 \pm 608$\\
\enddata
%\hline
\end{deluxetable*}

%% file: table2.tex
\begin{deluxetable*}{ccccc}
\tablecaption{Number of selected and reconstructed events for the $Z=2$ sample.\label{table:events_he}}
\tabletypesize{\scriptsize}
\tablewidth{0pt}
%\hline
%\hline
\tablehead{ \multicolumn{1}{c}{Kinetic energy} & \multicolumn{1}{c}{\hef\ events} & \multicolumn{1}{c}{\het\ events} &   %\multicolumn{1}{c}{\het\ contamination} &
                     \multicolumn{1}{c}{\hef\ extrapolated events} & \multicolumn{1}{c}{\het\ extrapolated events} \\
%\multicolumn{1}{c}{(GeV)} & \multicolumn{4}{c}{} }
\multicolumn{1}{c}{per nucleon (GeV/n)} & \multicolumn{2}{c}{} &  \multicolumn{1}{c}{at top of payload ($N_\text{ToP}$)} &  \multicolumn{1}{c}{at top of payload ($N_\text{ToP}$)} }

%\hline
\startdata
%$0.080 - 0.089 $ &  $9068 \pm 330$ & $396 \pm 63$ & $14173 \pm 522$ & $635 \pm 102$\\
%$0.089 - 0.100 $ &  $11055 \pm 355$ & $616 \pm 71$ & $17110 \pm 550$ & $984 \pm 116$\\
%$0.100 - 0.112 $ &  $13900 \pm 444$ & $872 \pm 86$ & $21417 \pm 684$ & $1386 \pm 141$\\
%$0.112 - 0.126 $ &  $16913 \pm 522$ & $1172 \pm 91$ & $25689 \pm 785$ & $1840 \pm 144$\\
$0.126 - 0.141 $ &  $21092 \pm 636$ & $1522 \pm 107$ & $32094 \pm 964$ & $2387 \pm 170$\\
$0.141 - 0.158 $ &  $24709 \pm 652$ & $1929 \pm 107$ & $37995 \pm 1015$ & $3023 \pm 170$\\
$0.158 - 0.177 $ &  $29506 \pm 695$ & $2403 \pm 131$ & $45235 \pm 1068$ & $3768 \pm 208$\\
$0.177 - 0.198 $ &  $35215 \pm 817$ & $2952 \pm 143$ & $54237 \pm 1243$ & $4645 \pm 226$\\
$0.198 - 0.222 $ &  $41078 \pm 897$ & $3583 \pm 162$ & $64104 \pm 1405$ & $5618 \pm 256$\\
$0.222 - 0.249 $ &  $47654 \pm 980$ & $4306 \pm 163$ & $74926 \pm 1545$ & $6864 \pm 264$\\
$0.249 - 0.279 $ &  $53935 \pm 1077$ & $5124 \pm 183$ & $84403 \pm 1640$ & $8203 \pm 295$\\
$0.279 - 0.312 $ &  $61541 \pm 1148$ & $6040 \pm 196$ & $98143 \pm 1851$ & $9853 \pm 325$\\
$0.312 - 0.350 $ &  $68608 \pm 1145$ & $7046 \pm 207$ & $109058 \pm 1837$ & $11645 \pm 345$\\
$0.350 - 0.392 $ &  $76090 \pm 1224$ & $8125 \pm 226$ & $121874 \pm 1971$ & $13737 \pm 388$\\
$0.392 - 0.439 $ &  $83651 \pm 1268$ & $9246 \pm 255$ & $135138 \pm 2058$ & $16080 \pm 453$\\
$0.439 - 0.492 $ &  $90552 \pm 1262$ & $10353 \pm 266$ & $143775 \pm 2017$ & $18371 \pm 471$\\
$0.492 - 0.551 $ &  $99140 \pm 1422$ & $11366 \pm 286$ & $159728 \pm 2310$ & $20577 \pm 521$\\
$0.551 - 0.618 $ &  $104636 \pm 1483$ & $12173 \pm 321$ & $167348 \pm 2397$ & $23111 \pm 612$\\
$0.618 - 0.692 $ &  $110854 \pm 1665$ & $13209 \pm 332$ & $181468 \pm 2771$ & $26133 \pm 662$\\
$0.692 - 0.776 $ &  $114097 \pm 1624$ & $13946 \pm 363$ & $186092 \pm 2665$ & $28768 \pm 762$\\
$0.776 - 0.870 $ &  $117080 \pm 1768$ & $13985 \pm 362$ & $193787 \pm 2935$ & $30778 \pm 811$\\
%$0.870 - 0.974 $ &  $118142 \pm 1850$ & $13172 \pm 257$ & $191384 \pm 3024$ & $30702 \pm 530$\\
%$0.974 - 1.092 $ &  $118707 \pm 1965$ & $11654 \pm 241$ & $197280 \pm 3312$ & $28930 \pm 545$\\
%$1.092 - 1.223 $ &  $117297 \pm 1986$ & $10254 \pm 178$ & $196550 \pm 3397$ & $28130 \pm 503$\\
\enddata
%\hline
\end{deluxetable*}

%% file: table3.tex
\begin{deluxetable*}{lcccc}
\tablecaption{Selection efficiencies, divided by detector.\label{table:eff}}
\tabletypesize{\scriptsize} 
\tablewidth{0pt}
%\hline
%\hline
\tablehead{ \multicolumn{1}{c}{} & \multicolumn{1}{c}{\prot} & \multicolumn{1}{c}{\deu} &   \multicolumn{1}{c}{\het} &
                     \multicolumn{1}{c}{\hef} }

%\hline
\startdata
ToF and AC selections     & $88.99\% \pm 0.13\%$ & $88.51\% \pm 0.18\%$ & $89.0\% \pm 0.4\%$ & $88.1\% \pm 0.4\%$\\
Tracker selections (2006) & $88.6\% \pm 0.2\%$   & $87.1\% \pm 0.3\%$   & $89.5\% \pm 0.4\%$ & $88.3\% \pm 0.4\%$\\
Tracker selections (2007) & $62.2\% \pm 0.3\%$   & $60.8\% \pm 0.4\%$   & $61.2\% \pm 0.6\%$ & $62.1\% \pm 0.7\%$\\
Tracker $dE/dx$ selection & $100\%$                    & $96.67\% \pm 0.13\%$ & $95.5\% \pm 0.2\%$                  & $100\%$ \\
\hline\hline
Total efficiency (2006)        & $78.8\% \pm 0.3\%$ & $74.5\% \pm 0.5\%$ & $76.1\% \pm 0.8\%$ & $77.8\% \pm 0.7\%$\\
Total efficiency (2007)        & $55.4\% \pm 0.5\%$ & $52.0\% \pm 0.5\%$ & $52.0\% \pm 0.9\%$ & $54.7\% \pm 1.2\%$\\
\enddata
\end{deluxetable*}

%% file: table4.tex
\begin{deluxetable*}{cccc}
\tablecaption{Hydrogen isotope fluxes and their ratio, errors are statistical and systematics respectively. \label{tab:hydrogen}}
\tabletypesize{\scriptsize}
\tablewidth{0pt}
\tablehead{ 
 \multicolumn{1}{c}{Kinetic energy} & \multicolumn{1}{c}{\prot\ flux} & \multicolumn{1}{c}{\deu\ flux} & \multicolumn{1}{c}{\deu / \prot} \\
 \multicolumn{1}{c}{at top of payload} & \multicolumn{3}{c}{} \\
 \multicolumn{1}{c}{(GeV n$^{-1}$)} & \multicolumn{1}{c}{(GeV n$^{-1}$ m$^2$ s sr)$^{-1}$} &  \multicolumn{1}{c}{(GeV n$^{-1}$ m$^2$ s sr)$^{-1}$} &
 }
\startdata
0.120 - 0.132 & $(9.86 \pm 0.15 \pm 0.31) \cdot 10^{2}$ & $(30.8 \pm 1.8 \pm 1.0)$ & $(3.12 \pm 0.19 \pm 0.20) \cdot 10^{-2}$ \\ 
0.132 - 0.144 & $(1.064 \pm 0.014 \pm 0.030) \cdot 10^{3}$ & $(33.0 \pm 1.6 \pm 0.9)$ & $(3.10 \pm 0.15 \pm 0.18) \cdot 10^{-2}$ \\ 
0.144 - 0.158 & $(1.136 \pm 0.013 \pm 0.029) \cdot 10^{3}$ & $(34.2 \pm 1.4 \pm 0.9)$ & $(3.01 \pm 0.13 \pm 0.16) \cdot 10^{-2}$ \\ 
0.158 - 0.173 & $(1.182 \pm 0.012 \pm 0.028) \cdot 10^{3}$ & $(35.1 \pm 1.4 \pm 0.8)$ & $(2.97 \pm 0.12 \pm 0.14) \cdot 10^{-2}$ \\ 
0.173 - 0.190 & $(1.233 \pm 0.012 \pm 0.027) \cdot 10^{3}$ & $(35.3 \pm 1.3 \pm 0.8)$ & $(2.86 \pm 0.11 \pm 0.13) \cdot 10^{-2}$ \\ 
0.190 - 0.208 & $(1.300 \pm 0.011 \pm 0.027) \cdot 10^{3}$ & $(36.6 \pm 1.3 \pm 0.8)$ & $(2.81 \pm 0.10 \pm 0.12) \cdot 10^{-2}$ \\ 
0.208 - 0.228 & $(1.351 \pm 0.010 \pm 0.027) \cdot 10^{3}$ & $(37.1 \pm 1.2 \pm 0.7)$ & $(2.74 \pm 0.09 \pm 0.11) \cdot 10^{-2}$ \\ 
0.228 - 0.250 & $(1.405 \pm 0.010 \pm 0.027) \cdot 10^{3}$ & $(36.3 \pm 1.1 \pm 0.7)$ & $(2.58 \pm 0.08 \pm 0.10) \cdot 10^{-2}$ \\ 
0.250 - 0.274 & $(1.453 \pm 0.009 \pm 0.027) \cdot 10^{3}$ & $(34.5 \pm 1.0 \pm 0.6)$ & $(2.37 \pm 0.07 \pm 0.09) \cdot 10^{-2}$ \\ 
0.274 - 0.300 & $(1.486 \pm 0.008 \pm 0.027) \cdot 10^{3}$ & $(33.9 \pm 0.9 \pm 0.6)$ & $(2.28 \pm 0.06 \pm 0.08) \cdot 10^{-2}$ \\ 
0.300 - 0.329 & $(1.514 \pm 0.008 \pm 0.027) \cdot 10^{3}$ & $(34.1 \pm 0.9 \pm 0.6)$ & $(2.25 \pm 0.06 \pm 0.08) \cdot 10^{-2}$ \\ 
0.329 - 0.361 & $(1.517 \pm 0.007 \pm 0.026) \cdot 10^{3}$ & $(33.6 \pm 0.9 \pm 0.6)$ & $(2.22 \pm 0.06 \pm 0.08) \cdot 10^{-2}$ \\ 
0.361 - 0.395 & $(1.520 \pm 0.007 \pm 0.026) \cdot 10^{3}$ & $(32.0 \pm 0.8 \pm 0.6)$ & $(2.10 \pm 0.06 \pm 0.07) \cdot 10^{-2}$ \\ 
0.395 - 0.433 & $(1.484 \pm 0.006 \pm 0.025) \cdot 10^{3}$ & $(31.4 \pm 0.8 \pm 0.6)$ & $(2.11 \pm 0.05 \pm 0.07) \cdot 10^{-2}$ \\ 
0.433 - 0.475 & $(1.480 \pm 0.006 \pm 0.025) \cdot 10^{3}$ & $(30.6 \pm 0.8 \pm 0.5)$ & $(2.07 \pm 0.05 \pm 0.07) \cdot 10^{-2}$ \\ 
0.475 - 0.520 & $(1.499 \pm 0.006 \pm 0.025) \cdot 10^{3}$ & $(29.9 \pm 0.8 \pm 0.5)$ & $(2.00 \pm 0.05 \pm 0.07) \cdot 10^{-2}$ \\ 
0.520 - 0.570 & $(1.429 \pm 0.005 \pm 0.024) \cdot 10^{3}$ & $(29.0 \pm 0.8 \pm 0.5)$ & $(2.03 \pm 0.06 \pm 0.07) \cdot 10^{-2}$ \\ 
\enddata
\end{deluxetable*}

%% file: table5.tex
%\begin{center}
%\begin{tabular}{ c  c  c  c }
%\hline
%\hline
% Kinetic energy & \hef\ flux & \het\ flux & \het / \hef \\
% at top of payload & & & \\
% (GeV n$^{-1}$) & (GeV n$^{-1}$ m$^2$ s sr)$^{-1}$ & (GeV n$^{-1}$ m$^2$ s sr)$^{-1}$ & \\
%\hline
\begin{deluxetable*}{cccc}
\tablecaption{Helium isotope fluxes and their ratio, errors are statistical and systematics respectively. \label{tab:helium}}
\tabletypesize{\scriptsize}
\tablewidth{0pt}
\tablehead{ 
 \multicolumn{1}{c}{Kinetic energy} & \multicolumn{1}{c}{\hef\ flux} & \multicolumn{1}{c}{\het\ flux} & \multicolumn{1}{c}{\het / \hef} \\
 \multicolumn{1}{c}{at top of payload} & \multicolumn{3}{c}{} \\
 \multicolumn{1}{c}{(GeV n$^{-1}$)} & \multicolumn{1}{c}{(GeV n$^{-1}$ m$^2$ s sr)$^{-1}$} &  \multicolumn{1}{c}{(GeV n$^{-1}$ m$^2$ s sr)$^{-1}$} &
 }
\startdata
0.126 - 0.141 & $(2.03 \pm 0.06 \pm 0.04) \cdot 10^{2}$ & $(17.1 \pm 1.8 \pm 0.4)$ & $(8.4 \pm 0.9 \pm 0.4) \cdot 10^{-2}$ \\ 
0.141 - 0.158 & $(2.20 \pm 0.06 \pm 0.04) \cdot 10^{2}$ & $(21.3 \pm 1.7 \pm 0.5)$ & $(9.7 \pm 0.8 \pm 0.4) \cdot 10^{-2}$ \\ 
0.158 - 0.177 & $(2.27 \pm 0.06 \pm 0.04) \cdot 10^{2}$ & $(23.4 \pm 1.5 \pm 0.5)$ & $(1.03 \pm 0.07 \pm 0.04) \cdot 10^{-1}$ \\ 
0.177 - 0.198 & $(2.33 \pm 0.06 \pm 0.04) \cdot 10^{2}$ & $(24.8 \pm 1.4 \pm 0.5)$ & $(1.06 \pm 0.06 \pm 0.04) \cdot 10^{-1}$ \\ 
0.198 - 0.222 & $(2.38 \pm 0.05 \pm 0.04) \cdot 10^{2}$ & $(25.4 \pm 1.2 \pm 0.5)$ & $(1.07 \pm 0.06 \pm 0.04) \cdot 10^{-1}$ \\ 
0.222 - 0.249 & $(2.42 \pm 0.05 \pm 0.04) \cdot 10^{2}$ & $(26.7 \pm 1.2 \pm 0.5)$ & $(1.10 \pm 0.05 \pm 0.04) \cdot 10^{-1}$ \\ 
0.249 - 0.279 & $(2.39 \pm 0.05 \pm 0.04) \cdot 10^{2}$ & $(27.3 \pm 1.0 \pm 0.5)$ & $(1.14 \pm 0.05 \pm 0.04) \cdot 10^{-1}$ \\ 
0.279 - 0.312 & $(2.37 \pm 0.04 \pm 0.04) \cdot 10^{2}$ & $(28.2 \pm 1.0 \pm 0.5)$ & $(1.19 \pm 0.05 \pm 0.04) \cdot 10^{-1}$ \\ 
0.312 - 0.350 & $(2.29 \pm 0.04 \pm 0.04) \cdot 10^{2}$ & $(28.4 \pm 0.9 \pm 0.5)$ & $(1.24 \pm 0.04 \pm 0.04) \cdot 10^{-1}$ \\ 
0.350 - 0.392 & $(2.22 \pm 0.04 \pm 0.04) \cdot 10^{2}$ & $(28.9 \pm 0.8 \pm 0.6)$ & $(1.30 \pm 0.04 \pm 0.05) \cdot 10^{-1}$ \\ 
0.392 - 0.439 & $(2.11 \pm 0.03 \pm 0.04) \cdot 10^{2}$ & $(29.2 \pm 0.8 \pm 0.6)$ & $(1.39 \pm 0.04 \pm 0.05) \cdot 10^{-1}$ \\ 
0.439 - 0.492 & $(1.93 \pm 0.03 \pm 0.03) \cdot 10^{2}$ & $(28.6 \pm 0.8 \pm 0.6)$ & $(1.49 \pm 0.04 \pm 0.06) \cdot 10^{-1}$ \\ 
0.492 - 0.551 & $(1.84 \pm 0.03 \pm 0.03) \cdot 10^{2}$ & $(27.3 \pm 0.7 \pm 0.6)$ & $(1.49 \pm 0.04 \pm 0.06) \cdot 10^{-1}$ \\ 
0.551 - 0.618 & $(1.72 \pm 0.02 \pm 0.03) \cdot 10^{2}$ & $(26.5 \pm 0.7 \pm 0.5)$ & $(1.55 \pm 0.05 \pm 0.06) \cdot 10^{-1}$ \\ 
0.618 - 0.692 & $(1.59 \pm 0.02 \pm 0.03) \cdot 10^{2}$ & $(25.4 \pm 0.7 \pm 0.6)$ & $(1.60 \pm 0.05 \pm 0.06) \cdot 10^{-1}$ \\ 
0.692 - 0.776 & $(1.44 \pm 0.02 \pm 0.03) \cdot 10^{2}$ & $(24.5 \pm 0.6 \pm 0.5)$ & $(1.70 \pm 0.05 \pm 0.07) \cdot 10^{-1}$ \\ 
0.776 - 0.870 & $(1.292 \pm 0.020 \pm 0.023) \cdot 10^{2}$ & $(23.0 \pm 0.6 \pm 0.5)$ & $(1.78 \pm 0.05 \pm 0.07) \cdot 10^{-1}$ \\ 
\enddata
\end{deluxetable*}
%\hline
%\end{tabular}
%\end{center}